\documentclass{PoS}

\usepackage{subcaption}

\def\be{\begin{equation}}
\def\ee{\end{equation}}
\def\bea{\begin{eqnarray}}
\def\eea{\end{eqnarray}}
\def\nn{\nonumber}
\newcommand\TwoFigBottom{-2}
\newcommand{\cg}{c_{ggh}}
\newcommand{\cgg}{c_{gghh}}
\newcommand{\ctt}{c_{tt}}
\newcommand{\ct}{c_{t}}

\newcommand{\chhh}{c_{hhh}}
\newcommand{\mhh}{m_{hh}}

\newcommand{\ftapprox}{FT$_{\mathrm{approx}}$}

\title{Higher order and top mass effects in Higgs boson pair production beyond the Standard Model}

\ShortTitle{Higher order and top mass effects in Higgs pair production beyond the SM}

\author{Gerhard Buchalla\\
        Ludwig-Maximilians-Universit{\"a}t M{\"u}nchen, Fakult\"at f\"ur Physik,
Arnold Sommerfeld Center for Theoretical Physics, 80333 M\"unchen, Germany\\
        E-mail: \email{gerhard.buchalla@physik.uni-muenchen.de}}

\author{Matteo Capozi\\
              Max Planck Institute for Physics,
       F\"ohringer Ring 6,
       80805 M{\"u}nchen, Germany\\
        E-mail: \email{mcapozi@mpp.mpg.de}}

\author{Alejandro Celis\\
        Ludwig-Maximilians-Universit{\"a}t M{\"u}nchen, Fakult\"at f\"ur Physik,
Arnold Sommerfeld Center for Theoretical Physics, 80333 M\"unchen, Germany\\
        E-mail: \email{alejandro.celis@physik.uni-muenchen.de}}

\author{\speaker{Gudrun Heinrich}\\
        Max Planck Institute for Physics,
       F\"ohringer Ring 6,
       80805 M{\"u}nchen, Germany\\
        E-mail: \email{gudrun@mpp.mpg.de}}

\author{Ludovic Scyboz\\
         Max Planck Institute for Physics,
       F\"ohringer Ring 6,
       80805 M{\"u}nchen, Germany\\
        E-mail: \email{scyboz@mpp.mpg.de}}

\abstract{We discuss the interplay between NLO QCD corrections and anomalous couplings in Higgs boson pair production via gluon fusion, 
within the framework of a non-linearly realised Effective Field Theory, described by the electroweak chiral Lagrangian.
We study how the NLO corrections with full top quark mass dependence affect the total cross sections as well as distributions in the Higgs boson pair invariant mass.      
For a large part of the parameter space, significant and non-homogeneous K-factors arise.     
}

\FullConference{Loops and Legs in Quantum Field Theory (LL2018)\\
		29 April 2018 - 04 May 2018\\
		St. Goar, Germany}

\begin{document}

\section{Introduction}

At the energy scales which colliders probe currently and in the nearer future, physics beyond the Standard Model (BSM) may manifest itself indirectly, for example via anomalous couplings in the Higgs sector. 
Some of the Higgs boson couplings, in particular the self-coupling, are still largely unconstrained and leave room for New Physics~\cite{Brooijmans:2018xbu}.
Assuming a New Physics scale $\Lambda$ in the TeV range or above,  the BSM effects can be parametrised in a model-independent way in an Effective Field Theory (EFT) framework, where we can distinguish two main categories, often called ``linear EFT'' and ``non-linear EFT''. 
While the linear EFTs~\cite{Buchmuller:1985jz,Grzadkowski:2010es}, 
also known as ``SMEFT''~\cite{Berthier:2015oma}, 
are formulated as power series in the dimensionful parameter $1/\Lambda$, the non-linear EFTs are organised by chiral dimensions
and therefore the formalism is also called ``Electroweak Chiral Lagrangian'' (EWChL) framework. 
Prominent features of this approach are  that the anomalous Higgs couplings
are singled out systematically as the dominant New Physics effects
in the electroweak sector, and that the Higgs field is an electroweak singlet. 
For more details we refer to Refs.~\cite{Buchalla:2018yce,Alonso:2012px,Buchalla:2013rka} and references therein.

Higgs boson pair production in gluon fusion is the most promising process to measure the Higgs boson self-coupling
and possibly other (effective) couplings involving more than one Higgs boson. 
In the Standard Model (SM), Higgs boson pair production has been calculated at leading order in Refs.~\cite{Eboli:1987dy,Glover:1987nx}.
As it is a loop-induced process, higher order calculations with full top quark mass dependence involve multi-scale two-loop integrals. 
Therefore, the NLO calculations until recently have been performed in the $m_t\to\infty$ limit~\cite{Dawson:1998py}, 
also called HTL or HEFT (``Higgs Effective Field Theory'')\footnote{Sometimes the electroweak chiral Lagrangian with a light Higgs
boson is also referred to as {\it Higgs Effective Field Theory (HEFT)\/} in the
literature. To avoid confusion, we will employ here the term 
{\it electroweak chiral Lagrangian\/} for the non-linear EFT of physics beyond
the SM, and reserve the expression {\it HEFT\/} for the heavy-top limit.}, 
and then rescaled by a factor $B_{FT}/B_{HEFT}$, where $B_{FT}$
denotes the leading order matrix element squared in the full theory.
This procedure is called ``Born-improved HEFT'' in the following.
Further, in Refs.~\cite{Frederix:2014hta,Maltoni:2014eza}, an approximation called
``\ftapprox'' was introduced, which contains the full top quark
mass dependence in the Born and real radiation parts, while the virtual part is
calculated in the Born-improved HEFT approximation.

The full NLO corrections, including the top quark mass
dependence also in the virtual two-loop amplitudes, have been
calculated in Ref.~\cite{Borowka:2016ehy}, based on a numerical evaluation of the multi-scale two-loop integrals with the program {\sc SecDec}~\cite{Borowka:2015mxa,Borowka:2017idc}.
Phenomenological studies at 14\,TeV and 100\,TeV, including variations of the Higgs boson self-coupling, have been presented in Ref.~\cite{Borowka:2016ypz}.  
The full NLO calculation was supplemented by NLL transverse momentum resummation in Ref.~\cite{Ferrera:2016prr}. 
It also has been matched to parton shower Monte Carlo programs~\cite{Heinrich:2017kxx,Jones:2017giv}.

The NNLO QCD corrections in the heavy-top limit have been computed in
Refs.~\cite{deFlorian:2013uza,deFlorian:2013jea,Grigo:2014jma,deFlorian:2016uhr},
and they have been supplemented by an expansion in $1/m_t^2$ in
Ref.~\cite{Grigo:2015dia} and by threshold resummation~\cite{Shao:2013bz,deFlorian:2015moa}.
In Ref.~\cite{Grazzini:2018bsd}, top quark mass effects have been incorporated in the NNLO HEFT calculation, 
including the full NLO result and combining
one-loop double-real corrections with full top mass dependence with suitably reweighted
real-virtual and double-virtual contributions evaluated in the large-$m_t$ approximation.
Very recently, threshold resummation on top of the latter result has been worked out in Ref.~\cite{deFlorian:2018tah}.


Within a non-linear EFT framework, higher order QCD corrections have been performed in the $m_t\to\infty$ limit.
The NLO QCD corrections have been calculated in Ref.~\cite{Grober:2015cwa}, including the case of CP-violating Higgs sectors~\cite{Grober:2017gut}. 
The NNLO QCD corrections in the $m_t\to\infty$ limit including dimension~6 operators have been presented in Ref.~\cite{deFlorian:2017qfk}.
These calculations found rather flat K-factors, which however could be an artefact of the $m_t\to\infty$ limit. 

Here we investigate whether this feature is preserved once the full top quark mass dependence is taken into account at NLO QCD, 
and quantify the  effects of  five operators  that can lead to deviations from the SM in the Higgs sector.

\section{Setup of the calculation}

\begin{figure}[h]
\begin{center}
\includegraphics[width=8cm]{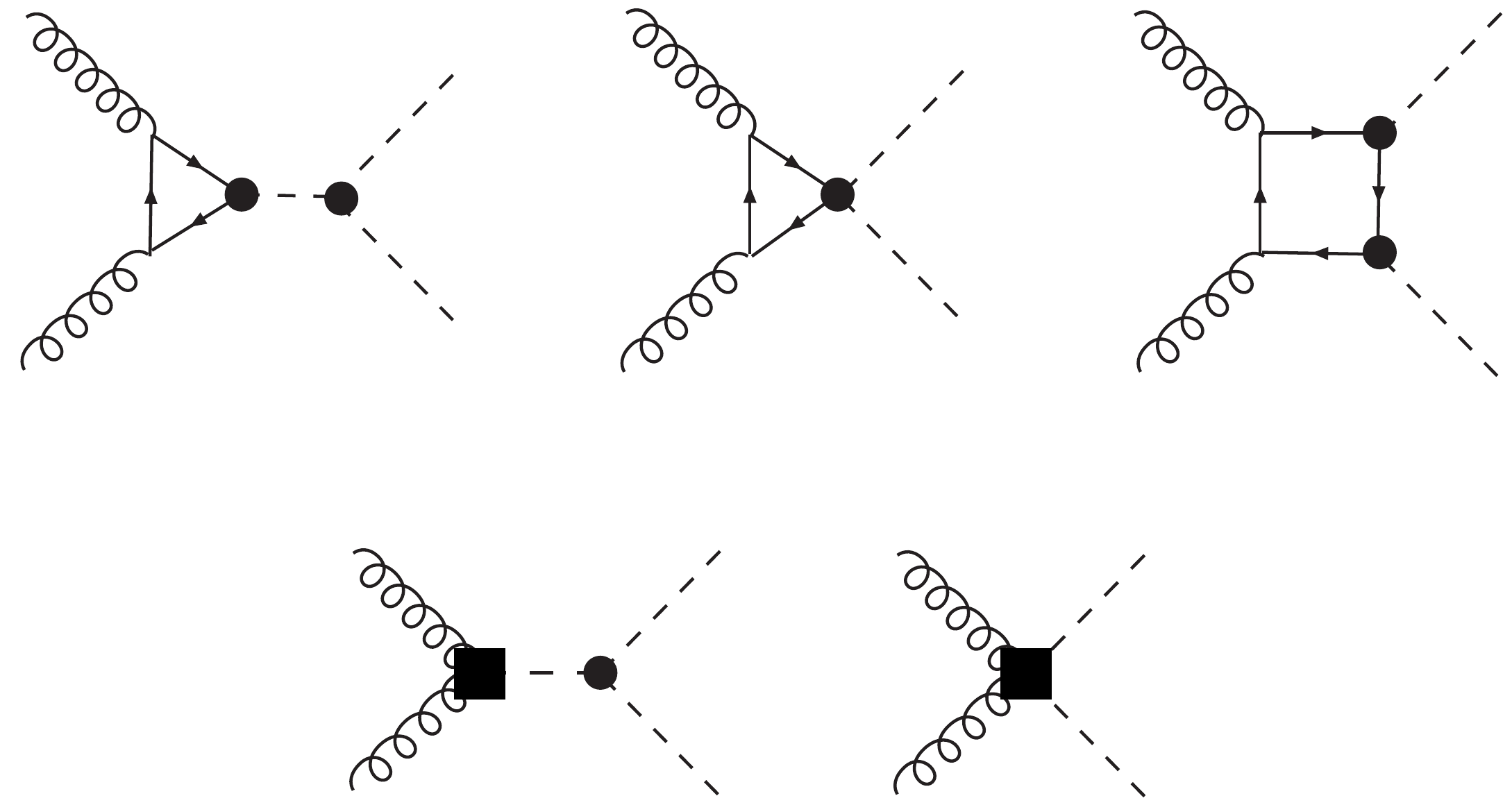}
\end{center}
\caption{Higgs-pair production in gluon fusion at leading order
in the chiral Lagrangian.}
\label{fig:hprocess}
\end{figure}

\begin{figure}[htb]
\begin{center}
\includegraphics[width=10cm]{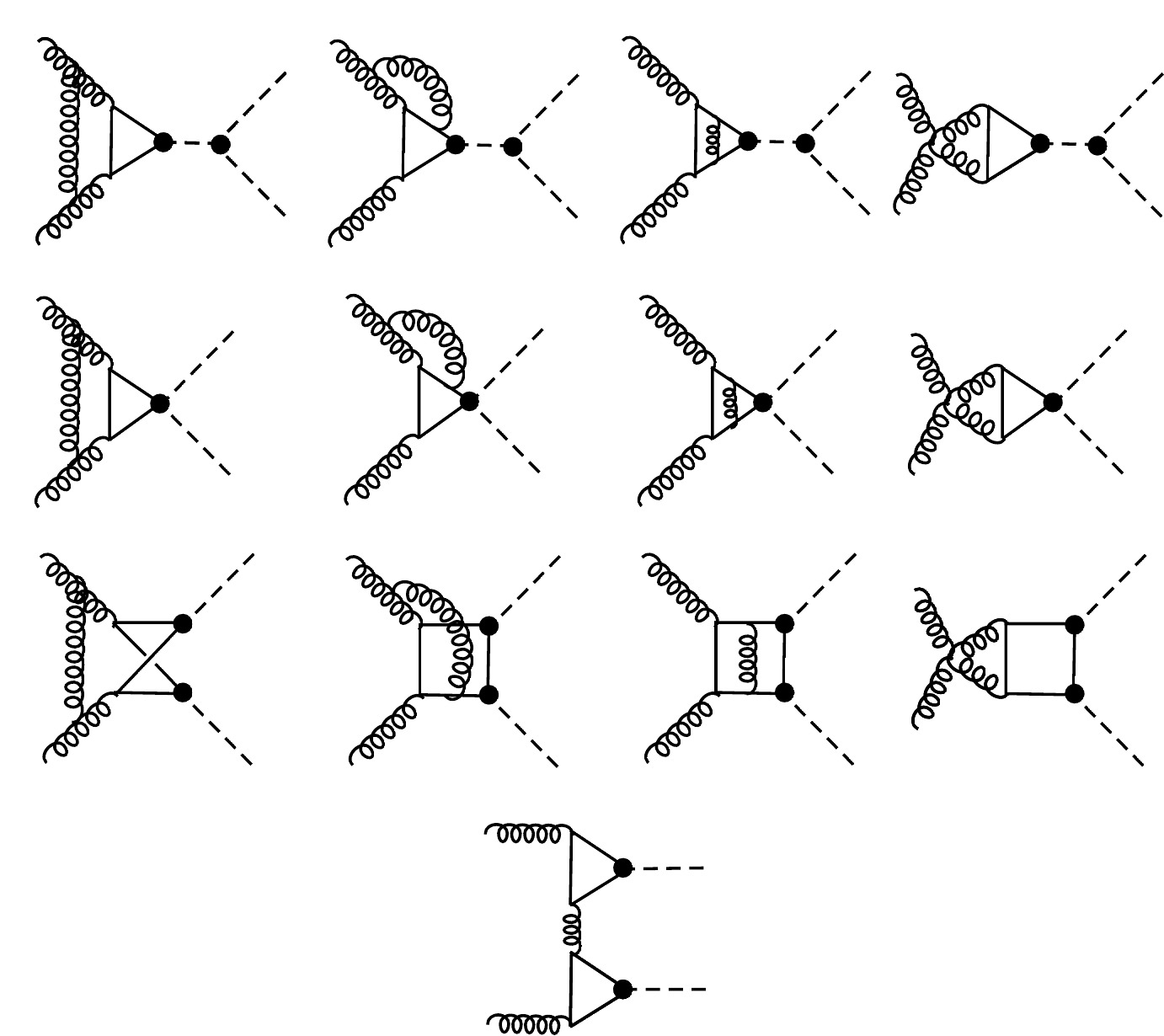}
\end{center}
\caption{Higgs-pair production in gluon fusion at NLO:
Examples for virtual two-loop diagrams at order $g^4_s$.}
\label{fig:hpnlov2}
\end{figure}

The terms in the effective Lagrangian relevant to our analysis are given by 
\be
{\cal L}\supset 
-m_t\left(c_t\frac{h}{\mathrm{v}}+c_{tt}\frac{h^2}{\mathrm{v}^2}\right)\,\bar{t}\,t -
c_{hhh} \frac{m_h^2}{2\mathrm{v}} \,h^3+\frac{\alpha_s}{8\pi} \left( c_{ggh} \frac{h}{\mathrm{v}}+
c_{gghh}\frac{h^2}{\mathrm{v}^2}  \right)\, G^a_{\mu \nu} G^{a,\mu \nu}\;.
\label{eq:ewchl}
\ee
To lowest order in the SM, $c_t=c_{hhh}=1$ and $c_{tt}=c_{ggh}=c_{gghh}=0$.
In principle, all couplings may have arbitrary values of ${\cal O}(1)$.
The leading-order diagrams are shown in Fig.~\ref{fig:hprocess}.
Examples of virtual NLO diagrams are shown in Fig.~\ref{fig:hpnlov2}.
For the two-loop part, we made use of the numerical results for the two-loop virtual diagrams in the SM~\cite{Borowka:2016ehy,Borowka:2016ypz} by dividing them into two classes: 
diagrams containing the Higgs-boson self-coupling (``triangle-type''), and diagrams without (``box-type''). 
The $t\bar{t}hh$ coupling generates new two-loop topologies, see e.g. the second line of Fig.~\ref{fig:hpnlov2}.
The results for these diagrams however can be obtained from the SM
triangle-type diagrams by omitting the $s$-channel Higgs boson propagator and multiplying with $\ctt/\chhh$. 
The other two-loop diagrams occurring in our calculation have the same topologies as in the SM and therefore can be obtained by rescaling of the couplings.

The real corrections consist of 5-point one-loop topologies with closed top quark loops as well as tree-level diagrams. 
Both classes of diagrams have been generated with {\sc GoSam}~\cite{Cullen:2011ac,Cullen:2014yla} in combination with a model file in {\sc ufo} format~\cite{Degrande:2011ua}, derived from our effective Lagrangian using {\sc FeynRules}~\cite{Alloul:2013bka}.
The various building blocks are assembled in a {\tt C++} program  and integrated over the phase space using 
the  {\sc Cuba} library~\cite{Hahn:2004fe}.

\section{Results}

All our results are for a centre-of-mass energy of $\sqrt{s}=14$\,TeV, computed using  $m_h=125$\,GeV, $m_t=173$\,GeV and the
PDF4LHC15{\tt\_}nlo{\tt\_}100{\tt\_}pdfas~\cite{Butterworth:2015oua}
parton distribution functions. 
The widths of the top quark (and the Higgs boson) have been set to zero.
Bottom quarks are treated as massless and therefore are
not included in the fermion loops.
The scale uncertainties are estimated by varying the factorisation scale $\mu_{F}$ and the
renormalisation scale $\mu_{R}$ 
around the central scale $\mu_0 =\mhh/2$, using the envelope of a 7-point scale variation.

\subsection{Quantifying the NLO corrections}

The total cross section can be written in terms of the 15 coefficients 
$A_1, \ldots, A_{15}$, at LO~\cite{Azatov:2015oxa,Carvalho:2015ttv} and in terms of 23 coefficients at NLO~\cite{Buchalla:2018yce}.
\bea
\label{eq:Acoeffs_all}
\sigma^{\rm{NLO}}/\sigma^{\rm{NLO}}_{SM}  &=& \quad  A_1\, c_t^4 + A_2 \, c_{tt}^2  + A_3\,  c_t^2 \chhh^2  + 
A_4 \, \cg^2 \chhh^2  + A_5\,  \cgg^2  + 
A_6\, c_{tt} c_t^2 + A_7\,  c_t^3 \chhh \nn\\
&& + A_8\,  c_{tt} c_t\, \chhh  + A_9\, c_{tt} \cg \chhh + A_{10}\, c_{tt} \cgg + 
A_{11}\,  c_t^2 \cg \chhh + A_{12}\, c_t^2 \cgg \nn\\
&& + A_{13}\, c_t \chhh^2 \cg  + A_{14}\, c_t \chhh \cgg +
A_{15}\, \cg \chhh \cgg \nn\\ 
&& + A_{16}\, c^3_t \cg + A_{17}\,  c_t c_{tt} \cg 
+ A_{18}\, c_t \cg^2 \chhh + A_{19}\, c_t \cg \cgg 
\nn\\
&&+ A_{20}\,  c_t^2 \cg^2 + A_{21}\, c_{tt} \cg^2 
+ A_{22}\, \cg^3 \chhh + A_{23}\, \cg^2 \cgg\,.
\eea
Based on our results for $A_1,\ldots, A_{23}$, we produced heat maps for the ratio $\sigma/\sigma_{SM}$, 
varying two of the five parameters, while for the fixed parameters the SM values are used, along with
$\sigma_{SM}^{\rm{LO}}=19.85$\,fb,
$\sigma_{SM}^{\rm{NLO}}=32.95$\,fb.
The couplings are varied in a range which seems reasonable when taking into account the current constraints on the 
Higgs coupling measurements~\cite{Khachatryan:2016vau}, 
as well as recent limits on the di-Higgs production cross section~\cite{Sirunyan:2018iwt,Aaboud:2018ftw}.

\begin{figure}[htb]
\begin{center}
\begin{subfigure}{0.495\textwidth}
   \includegraphics[width=7.5cm]{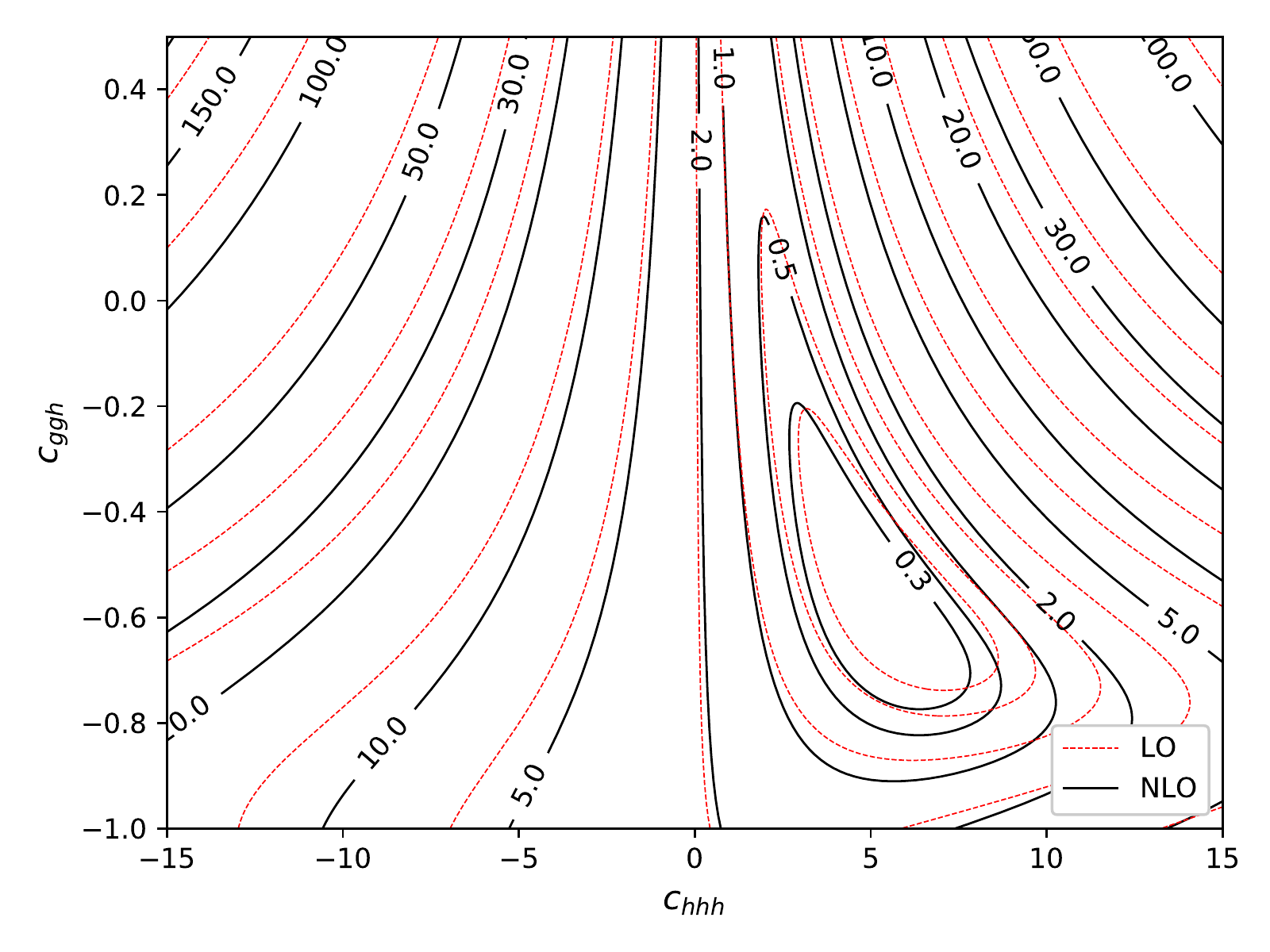}    
   \caption{\label{fig:chhh_cg}}
 \end{subfigure}
  \hfill
  \begin{subfigure}{0.495\textwidth}
\includegraphics[width=7.5cm]{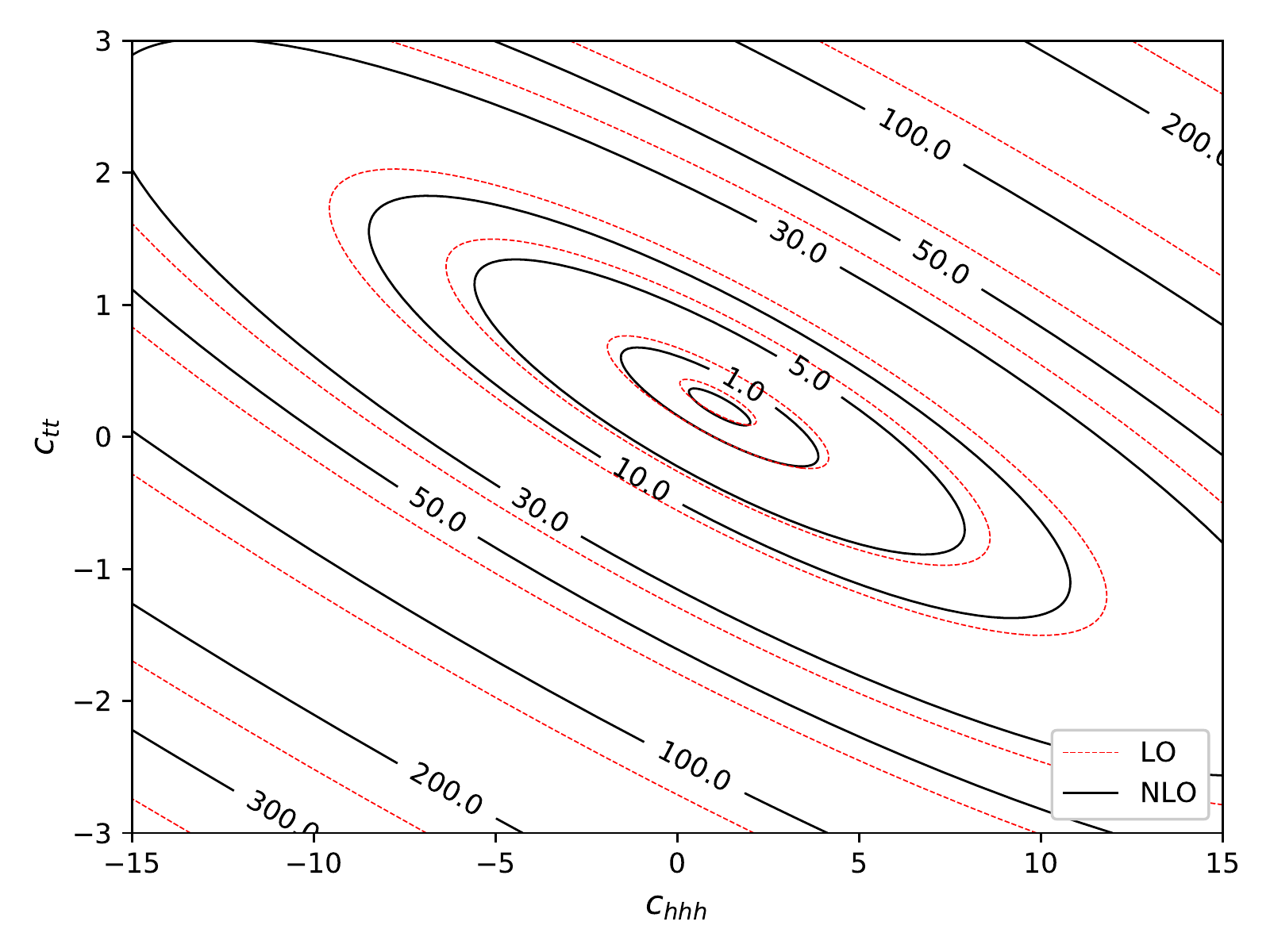}
  \caption{\label{fig:chhh_ctt}}
 \end{subfigure}
\end{center}
\caption{(a) $\cg$ versus $\chhh$ and (b) $\ctt$ versus $\chhh$.}
\label{fig:chhh_cg_ctt}
\end{figure}

\begin{figure}[htb]
\begin{center}
\includegraphics[width=8.5cm]{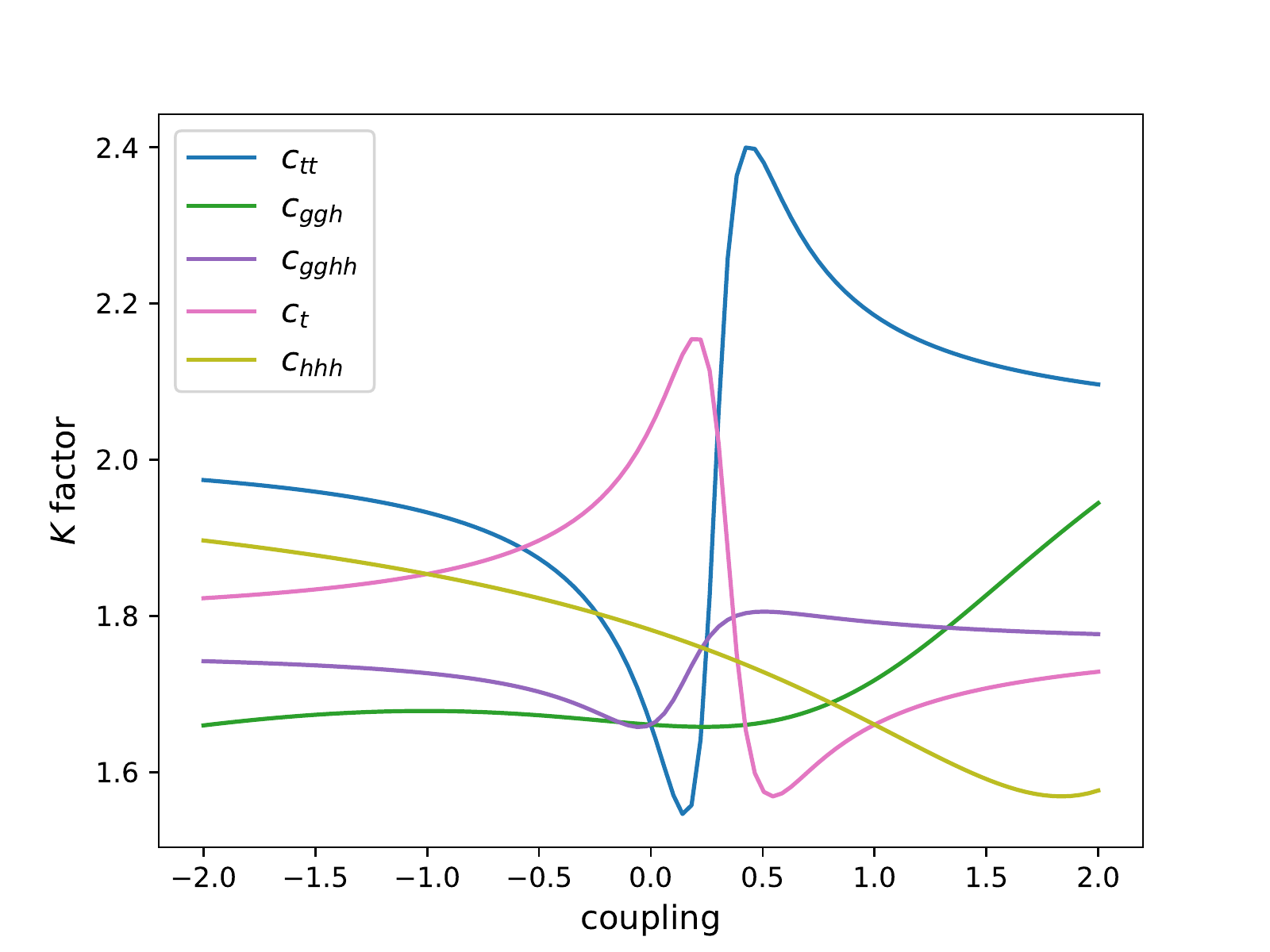}
\end{center}
\caption{K-factors for the total NLO cross section as a function of the
  different couplings.}
\label{fig:project_ctt}
\end{figure}

Fig.~\ref{fig:chhh_cg_ctt} shows the ratio to the SM total cross section both at LO and at NLO for variations of the triple Higgs coupling $\chhh$ in combination with $\cg$ and $\ctt$. 
We observe that the deviations from the SM cross section as well as the effects of the NLO corrections can be
substantial. Further we see a rapid variation of the cross section when changing $\ctt$ or $\chhh$, while it is less sensitive to changes of $\cg$. 
In Fig.~\ref{fig:project_ctt} we show the K-factors as a function of
the five coupling parameters (the fixed ones having SM values). 
It shows that the K-factors show a rather strong dependence on these
parameters, which was not the case in the $m_t\to \infty$ limit~\cite{Grober:2015cwa,deFlorian:2017qfk}.

\subsection{Total cross sections and distributions at benchmark points}

In Table~\ref{sigmatot} we show results for two benchmark points,  labeled as benchmarks 6 and 9 in Ref.~\cite{Carvalho:2015ttv}, 
where (in our conventions) the values for benchmark point 6 are $\chhh=2.4,\ct= 1, \ctt=0, \cg=2/15,\cgg=1/15$, and the ones for 
benchmark point 9 are $\chhh=1,\ct= 1, \ctt=1, \cg=-0.4,\cgg=-0.2$.
The corresponding total cross sections are shown in Table~\ref{sigmatot}.
Results for  10 more benchmark points can be found in Ref.~\cite{Buchalla:2018yce}.

\begin{table}[htb]
\begin{center}
\begin{tabular}{|c|c|c|c|c|c|}
\hline
Benchmark & $\sigma_{NLO}$ [fb] & K-factor & scale uncert.  [\%] &
stat. uncert. [\%]  &$\frac{\sigma_{NLO}}{\sigma_{NLO,SM}}$ \\ 
\hline
$   B_6$ & 24.69 &  1.89& ${+2\atop -11}$  & 2.1 & 0.7495  \\
\hline 
$   B_9$ & 146.00 & 2.30& ${+22\atop -16}$ & 0.31 & 4.431 \\
\hline 
$   SM $ & 32.95 & 1.66 & ${+14\atop -13}$ & 0.1 & 1\\
\hline
\end{tabular}
\end{center}
\caption{Total cross sections at NLO (second column) including the K-factor (third column), scale
  uncertainties (4th column) and statistical uncertainties (5th
  column) and the ratio to the SM total NLO cross section (6th column).\label{sigmatot}}
\end{table}
%

\vspace*{3mm}

Now we consider the Higgs boson pair invariant mass distribution for the two benchmark points. 
The full NLO result is shown in red,  the two approximations ``Born-improved NLO HEFT'' (purple) and \ftapprox{} (green) are also shown.
The leading order BSM results are shown in yellow,  the SM results are shown in blue (NLO) and black (LO).
The lower ratio plot shows the ratio of the two approximate results to the full NLO result.
The upper ratio plot shows the differential BSM K-factor, i.e. NLO$_{\rm{BSM}}$/LO$_{\rm{BSM}}$, both evaluated with the same PDFs.
\begin{figure}[htb]
  \centering
  \begin{subfigure}{0.495\textwidth}
    \includegraphics[width=\textwidth]{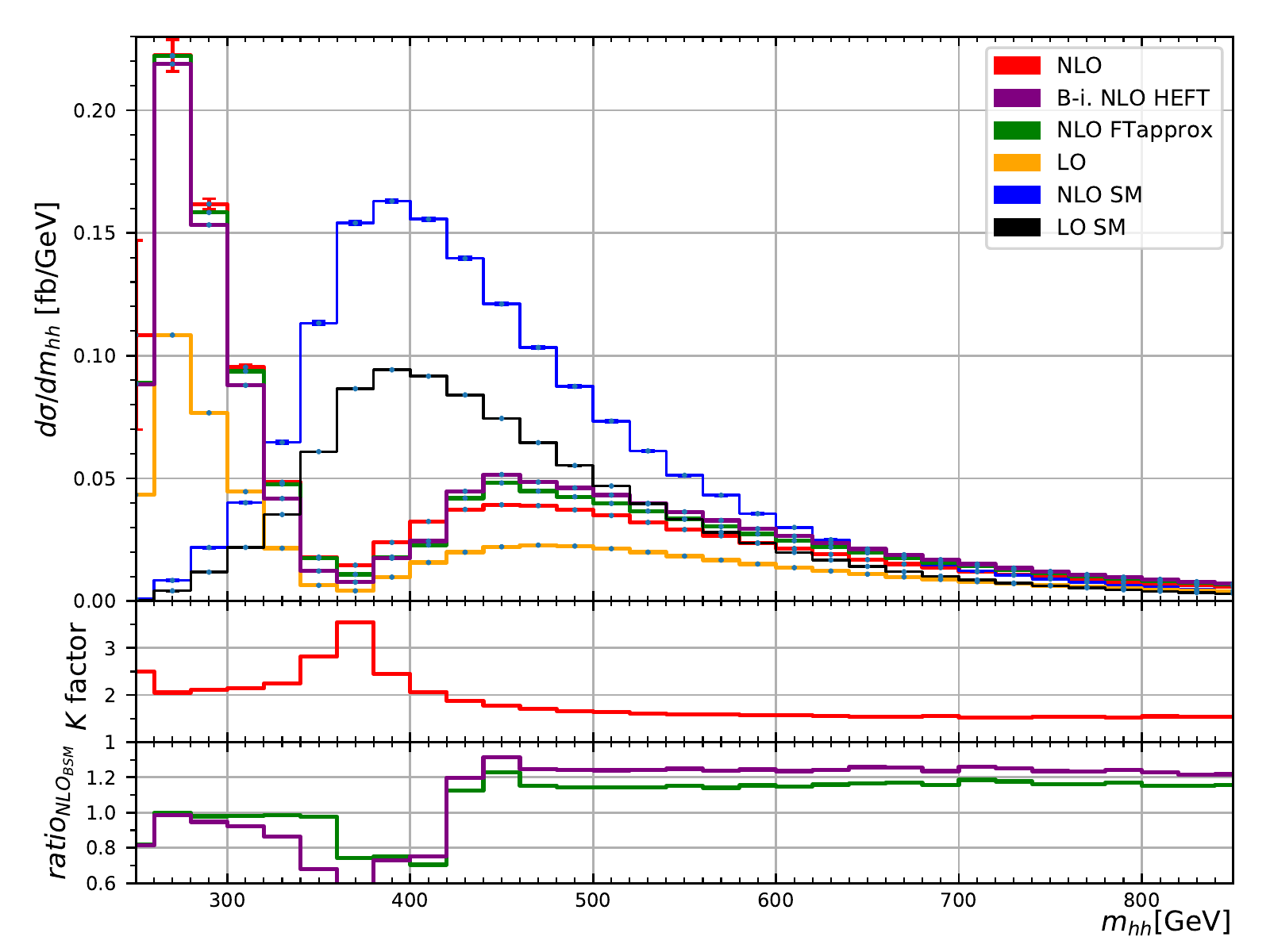}
    \vspace{\TwoFigBottom em}
    \caption{\label{fig:B6_mhh}}
  \end{subfigure}
  \hfill
  \begin{subfigure}{0.495\textwidth}
    \includegraphics[width=\textwidth]{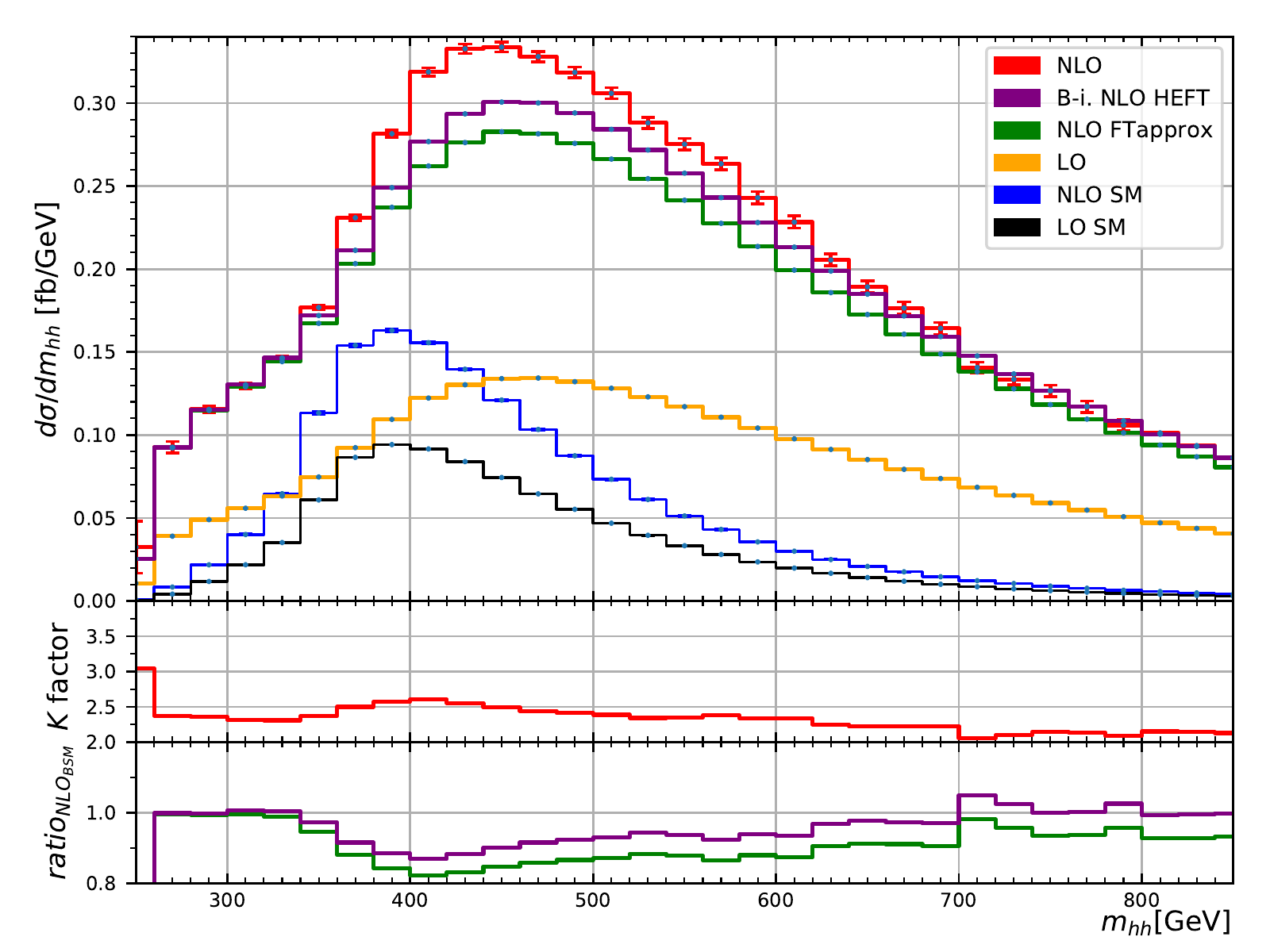}
    \vspace{\TwoFigBottom em}
    \caption{\label{fig:B9_mhh}}
  \end{subfigure}
\caption{Higgs boson pair invariant mass distribution for (a): Benchmark point 6 , $\chhh=2.4,\ct= 1, \ctt=0, \cg=2/15,\cgg=1/15$, 
          and (b) benchmark point 9, $\chhh=1,\ct= 1, \ctt=1, \cg=-0.4,\cgg=-0.2$.}
\label{fig:benchmarks}
\end{figure}

Benchmark point 6 shows a dip where the SM $\mhh$ distribution has a peak, 
related to the fact that the LO HEFT amplitude exactly vanishes at $\mhh=429$\,GeV. 
In addition it shows a large enhancement of the low $\mhh$ region due to the value $\chhh=2.4$.
Note  that this value for $\chhh$ is very close to the point where the 
total cross section as a function of $\chhh$ goes through a minimum if all other couplings are kept SM-like.
Even though the values for $\ct$ and $\ctt$ are the same as in the SM, 
the shape of the $\mhh$ distribution for benchmark 6 is very different from the SM one and therefore would be a very characteristic sign of anomalous couplings.

For benchmark point 9, the values for $\chhh$ and $\ct$ are as in the SM, however the total cross section is more than four times larger.
 The enhancement is particularly pronounced in the tails of the distributions, which can be attributed mainly to the rather large absolute values of $\cg$ and $\cgg$, in combination with a non-zero value of $\ctt$.

\section{Conclusions}

We have presented a calculation of the NLO QCD corrections with full $m_t$ dependence to Higgs boson
 pair production within the framework of a non-linearly realised
 Effective Field Theory (Electroweak Chiral Lagrangian) in the Higgs sector.
This framework,  applied to $gg\to hh$, allows us to focus on
 five anomalous Higgs boson  couplings, $\chhh,\ct,\ctt,\cg$ and $\cgg$.

Our calculation is based on a numerical evaluation of two-loop multi-scale integrals which so far are not accessible analytically.
In particular, the results for two-loop integrals involving an effective $hh\,t\bar{t}$ contact interaction, 
parametrised by $\ctt$, allowed us to study for the first time the effect of such an anomalous coupling including full NLO QCD corrections. 
We found that the cross sections are quite sensitive to variations of $\ctt$, while variations of the  Higgs-gluon effective interactions $\cg$ and $\cgg$
have a weaker effect on the total cross sections. 
For the considered benchmark points, the NLO K-factors are of the order of two for the total cross sections, 
however they can vary by up to $\pm 20\%$  as the anomalous couplings are varied. The differential K-factors for the $\mhh$ distribution show even stronger variations, in particular around the threshold region $\mhh\sim 2\,m_t$. 
This emphasises the importance of including the NLO QCD corrections with full top quark mass dependence in studies of anomalous couplings in the Higgs sector.

\bibliography{refsLL18}
\bibliographystyle{JHEP}

\end{document}